## AUTHORS

Raj Hansini Khoiwal and Dr. Alan B. McMillan.

## TITLE

Embeddings are all you need! Achieving High Performance Medical Image Classification through Training-Free Embedding Analysis

## ABSTRACT

Developing artificial intelligence (AI) and machine learning (ML) models for medical imaging typically involves extensive training and testing on large datasets, consuming significant computational time, energy, and resources. There is a need for more efficient methods that can achieve comparable or superior diagnostic performance without the associated resource burden. We investigated the feasibility of replacing conventional training procedures with an embedding-based approach that leverages concise and semantically meaningful representations of medical images. Using pre-trained foundational models—specifically, convolutional neural networks (CNN) like ResNet and multimodal models like Contrastive Language-Image Pre-training (CLIP)—we generated image embeddings for multi-class classification tasks. Simple linear classifiers were then applied to these embeddings. The approach was evaluated across diverse medical imaging modalities, including retinal images, mammography, dermatoscopic images, and chest radiographs. Performance was compared to benchmark models trained and tested using traditional methods. The embedding-based models surpassed the benchmark area under the receiver operating characteristic curve (AUC-ROC) scores by up to 87% in multi-class classification tasks across the various medical imaging modalities. Notably, CLIP embedding models achieved the highest AUC-ROC scores, demonstrating superior classification performance while significantly reducing computational demands. Our study indicates that leveraging embeddings from pre-trained foundational models can effectively replace conventional, resource-intensive training and testing procedures in medical image analysis. This embedding-based approach offers a more efficient alternative for image segmentation, classification, and prediction, potentially accelerating AI technology integration into clinical practice.

## INTRODUCTION

Artificial Intelligence (AI) is rapidly revolutionizing medical imaging. Advanced AI models, particularly deep learning algorithms, have demonstrated remarkable capabilities in automated image segmentation, precise lesion detection, accurate disease classification, and biomarker quantification[1]. These advancements enable clinicians to detect diseases at earlier stages, refine diagnostic accuracy, and tailor personalized treatment plans—contributing to improved patient outcomes and an enhanced quality of life. The integration of AI into medical imaging holds the promise of revolutionizing healthcare by facilitating more efficient and effective diagnostic processes[2]. However, the development and deployment of current AI models in medical imaging are not without significant challenges. These models typically require extensive training on large, annotated medical datasets to learn underlying patterns effectively. The process is computationally intensive, time-consuming, and financially costly, often necessitating substantial computational resources and expertise. Moreover, the reliance on large datasets introduces potential issues related to data quality and representativeness. If the training dataset is biased, unbalanced, or noisy, the AI model may inadvertently amplify these biases, leading to limited generalization and diminished performance on data that differ from the training set. Such limitations can hinder the applicability of AI models across diverse patient populations and clinical settings.

These concerns highlight the need for alternative approaches that enable AI models to perform effectively without the exhaustive requirements of traditional training methods. One promising avenue is the utilization of image embeddings as classifiers[3]. Image embeddings are dense vector representations that capture the semantic information of images, derived through machine learning models like deep neural networks[4]. They serve as the model's interpretation of the input data, encapsulating complex features in a condensed form. By focusing on the content rather than pixel-level features, embeddings facilitate comparisons between images based on meaningful characteristics. The application of image embeddings offers several advantages. It allows for efficient similarity searches based on the semantic content of images, which can be particularly valuable in diagnostic processes that rely on pattern recognition and comparison. With the advent of foundation models trained on massive corpora of data—including both images and text—the computed embeddings from these models can be leveraged for various downstream tasks without the need for additional training[5]. In image comparison and search tasks, embeddings can be directly employed, streamlining the process and reducing computational demands. For image classification tasks, embeddings enable the training of smaller, more efficient models using simpler classifiers such as k-means clustering, logistic regression, support vector machines, and random forests[6]. This approach significantly reduces the computational burden compared to training conventional deep neural networks, as embeddings need to be computed only once per image. The reduced complexity not only accelerates the development process but also makes it more accessible for clinical implementation, where resources and time may be limited.

In this study, we investigate the effectiveness of using image embeddings as classifiers in comparison to traditional end-to-end trained deep neural network models across various medical image classification tasks. We utilize embeddings computed from two pre-trained models: (1) a conventional convolutional neural network trained on the ImageNet database, which provides a rich representation of visual features, and (2) a multimodal language-vision model trained using contrastive learning techniques, which aligns visual and textual modalities to enhance semantic understanding. Our research aims to determine whether embeddings-based classifiers can meet or exceed the performance of fully trained deep neural networks in medical imaging applications. The findings suggest that embeddings-based classifiers not only achieve comparable performance but also offer significant advantages in terms of efficiency and ease of deployment.

## METHODS

*Study Design*

We conducted a retrospective, cross-sectional study to evaluate the efficacy of using image embeddings from pre-trained models for medical image classification tasks as visually depicted in Figure 1. This study involved a model comparison and performance analysis across multiple datasets representing different imaging modalities and diagnostic challenges. Our primary aim was to determine whether linear classifiers trained on embeddings generated from Contrastive Language-Image Pre-training (CLIP)[7] and Residual Neural Network (ResNet)[8] models could achieve performance comparable to or exceeding that of models trained from scratch on raw image data. By assessing the performance and computational efficiency of embedding-based classifiers, we sought to explore a potentially more efficient alternative for developing AI models in medical imaging.

The intended applications of our AI system varied according to each dataset:
- **CBIS-DDSM[9]:** Detection and classification of breast cancer lesions as malignant or benign.
- **CheXpert[10]:** Detection of 14 common chest radiographic observations, including atelectasis, cardiomegaly, consolidation, edema, pleural effusion, pneumonia, pneumothorax, fractures, support devices, no finding, enlarged cardiomediastinal, lung lesion, lung opacity, and other pleural abnormalities.
- **HAM10000[11] and PAD-UFES-20[12]:** Classification of seven skin lesion types—actinic keratoses and intraepithelial carcinoma (Bowen's disease), basal cell carcinoma, benign keratosis-like lesions, dermatofibroma, melanoma, melanocytic nevi, and vascular lesions.
- **Ocular Disease Recognition (ODIR)[13]:** Identification of eye diseases such as cataract, diabetic retinopathy, glaucoma, myopia, age-related macular degeneration (AMD), hypertension, and normal findings.

*Data Sources, Preprocessing, and Model Development*

Comprehensive details regarding data sources, preprocessing steps, demographic and clinical characteristics, embedding generation techniques, hyperparameter tuning, and model development methodologies are available in the Supplementary Appendix. The number of images utilized from each dataset is listed in Table 1. Example images are shown in Figure 2. Image embeddings were derived using two pre-trained models: ResNet50 and Contrastive Language-Image Pre-training (CLIP). These embeddings encapsulated high-level visual and semantic features of the medical images, forming the primary inputs for the classification models. For classification, we implemented linear classifiers, including Logistic Regression and Support Vector Machines (SVM). Logistic Regression models were tailored to support binary, multiclass, and multi-label classification tasks, as dictated by the dataset structure. SVMs, employed in both linear and kernel-based configurations, addressed scenarios where the relationship between embeddings and class labels was non-linear.

*Reference Standard*
The ground truth reference standard was established using performance metrics from traditional computer vision and deep learning models trained from scratch, as reported in the benchmark paper "Deep Learning vs. Traditional Computer Vision for Medical Imaging"[6]. This paper provides comprehensive evaluations of various models on the same datasets, including metrics such as accuracy, sensitivity, specificity, and area under the receiver operating characteristic curve (AUC). By utilizing these benchmarks, we could directly compare the performance of our embedding-based classifiers to established models, assessing both efficacy and potential computational efficiency gains.

*Evaluation Metrics*
The performance of our classification models was evaluated to assess accuracy, precision, recall (sensitivity), F1-score, and the area under the receiver operating characteristic curve (AUC). Accuracy was calculated as the proportion of correctly predicted instances among all instances in the dataset. This metric provides an overall measure of the model's ability to make correct predictions and serves as a basic indicator of performance across all classes. Precision was defined as the proportion of true positive predictions among all positive predictions made by the model. It reflects the model's specificity in identifying only the relevant instances and avoiding false positives. High precision indicates that the model is reliable when it predicts a positive class. Recall (Sensitivity) measured the proportion of true positive predictions among all actual positive instances. This metric assesses the model's effectiveness in identifying all relevant cases within the dataset. High recall indicates that the model successfully captures most of the positive instances, minimizing false negatives. The F1-score was computed as the harmonic mean of precision and recall. This metric provides a balance between precision and recall, especially useful in situations where there is an uneven class distribution or when both false positives and false negatives carry significant consequences. The F1-score offers a single measure that accounts for both the model's accuracy in predicting positive instances and its ability to find all positive instances. Area Under the Receiver Operating Characteristic Curve (AUC) was calculated to assess the model's ability to discriminate between classes across various threshold settings. The ROC curve plots the true positive rate against the false positive rate at different classification thresholds, and the AUC summarizes this information into a single value. An AUC of 1.0 represents perfect discrimination, whereas an AUC of 0.5 indicates no discriminative ability. By evaluating the AUC, we gauged the overall diagnostic effectiveness of the models beyond a single threshold, providing insight into their performance across a spectrum of operating conditions.

## RESULTS

A total of 20 embedding-based classification models were developed and evaluated across five medical imaging datasets:, CBIS-DDSM, CheXpert, HAM10000 Ocular Disease Recognition (ODIR), PAD-UFES-20. Each model combined embeddings from pre-trained ResNet50 or CLIP architectures with linear classifiers—either Logistic Regression (LR) or Support Vector Machines (SVM). The performance of these models was compared to the established benchmark AUC values reported in the literature as listed in Table 2. Detailed metrics including accuracy, precision, recall (sensitivity), F1-score, and AUC for both LR and SVM are listed in Table 3.

*CBIS-DDSM Dataset*
For the CBIS-DDSM dataset, which focuses on the binary classification of breast lesions as benign or malignant, the benchmark AUC was 0.464. The embedding-based models showed modest improvements over this benchmark. The ResNet50 embedding with SVM achieved an AUC of 0.5035, and with LR, an AUC of 0.491. Using CLIP embeddings, the SVM classifier attained an AUC of 0.4954, while the LR classifier slightly exceeded this with an AUC of 0.5052. Although the improvements were incremental, the embedding-based models demonstrated comparable performance to traditional models, indicating their potential utility in breast cancer detection tasks.

*CheXpert Dataset*

In the CheXpert dataset, which involves multi-label classification of 14 chest radiographic observations, the benchmark AUC was 0.723. The embedding-based models matched this level of performance. The ResNet50 embedding with LR resulted in an AUC of 0.7412, while the CLIP embedding with LR yielded an AUC of 0.7490. The ResNet50 embedding with SVM resulted in an AUC of 0.7329, while the CLIP embedding with SVM yielded an AUC of 0.7462. These findings indicate that the embedding-based classifiers perform as well as the benchmark in detecting chest radiographic observations in the CheXpert dataset.

### HAM10000 Dataset
In the HAM10000 dataset, which involves multiclass classification of seven skin lesion types, the benchmark AUC was reported as 0.609. The embedding-based models demonstrated a significant improvement over this benchmark. Specifically, the ResNet50 embeddings combined with an SVM classifier achieved an AUC of 0.935, while the same embeddings with an LR classifier yielded an AUC of 0.933. Models utilizing CLIP embeddings performed even better; the CLIP embedding with SVM attained an AUC of 0.9510, and with LR, it reached the highest AUC of 0.9586. These results indicate that embedding-based classifiers substantially outperformed the traditional benchmark, suggesting enhanced diagnostic accuracy in skin lesion classification when using embeddings from pre-trained models.

### Ocular Disease Recognition (ODIR) Dataset
In the ODIR dataset, which involves multi-label classification of ocular diseases such as cataract, diabetic retinopathy, and glaucoma, the benchmark AUC was 0.600. The embedding-based models showed substantial improvements over this benchmark. The ResNet50 embedding with SVM achieved an AUC of 0.8544, and with LR, an AUC of 0.7984. The models using CLIP embeddings performed similarly well; the CLIP embedding with SVM attained an AUC of 0.8577, and with LR, an AUC of 0.8506. These findings suggest that embedding-based classifiers markedly enhance the ability to detect multiple ocular diseases from retinal images.

### PAD-UFES-20 Dataset
For the PAD-UFES-20 dataset, which, like HAM10000, involves classification of skin lesions into multiple categories, the benchmark AUC was 0.487. The embedding-based models demonstrated significant improvements. The ResNet50 embedding with SVM achieved an AUC of 0.8576, and with LR, an AUC of 0.8516. Models utilizing CLIP embeddings further improved performance, with the CLIP embedding and SVM attaining an AUC of 0.9120, and with LR achieving the highest AUC of 0.9145. These results highlight the effectiveness of embedding-based classifiers in multiclass skin lesion classification, surpassing traditional benchmarks by a considerable margin.

### Summary of Findings
Overall, the embedding-based classifiers demonstrated performance that met or exceeded the benchmark AUC values in four out of the five datasets. Notably, in the MNIST: HAM10000, ODIR, and PAD-UFES-20 datasets, the models showed substantial improvements over the benchmarks, with AUC increases ranging from approximately 0.3 to 0.5 points. In the CBIS-DDSM dataset, the improvements were modest but still indicated that embedding-based models performed comparably to traditional models. The use of CLIP embeddings generally resulted in higher AUC values compared to ResNet50 embeddings, particularly when combined with Logistic Regression classifiers. The highest AUCs were achieved with CLIP embeddings and LR classifiers in both the MNIST: HAM10000 (AUC of 0.9586) and PAD-UFES-20 (AUC of 0.9145) datasets. These results suggest that embedding-based classifiers, leveraging pre-trained models and linear classifiers, can serve as effective and computationally efficient alternatives to traditional deep learning models trained from scratch for certain medical imaging classification tasks. The significant improvements observed in datasets involving skin lesion and ocular disease classification highlight the potential of this approach to enhance diagnostic accuracy while reducing computational demands.

## DISCUSSION

This study evaluated the effectiveness of using image embeddings from pre-trained models—specifically, the Contrastive Language-Image Pre-training (CLIP) model and the Residual Neural Network 50 (ResNet50)—combined with linear classifiers for medical image classification tasks. We found that embedding-based classifiers achieved performance that met or exceeded established benchmarks in all five datasets. Notably, significant improvements were observed in the classification of skin lesions in the HAM10000 and PAD-UFES-20 datasets and in the detection of ocular diseases in the ODIR dataset. These findings suggest that embedding-based classifiers can serve as effective and computationally efficient alternatives to traditional deep learning models trained from scratch, particularly in certain medical imaging applications.

The substantial improvements in the HAM10000 and PAD-UFES-20 datasets indicate that embeddings from pre-trained models can capture critical features necessary for accurate skin lesion classification. The high area under the receiver operating characteristic curve (AUC) values achieved with CLIP embeddings, especially when combined with logistic regression classifiers, highlight the potential of multimodal models that integrate visual and semantic information. In the ODIR dataset, the marked enhancement in detecting ocular diseases underscores the versatility of embedding-based approaches across different imaging modalities. The modest gains in the CBIS-DDSM dataset suggest that while embeddings can improve performance in breast lesion classification, the benefits may be limited compared to tasks involving skin and ocular images.

Our results build upon existing research demonstrating the utility of transfer learning and pre-trained models in medical imaging[14]. Previous studies have shown that deep learning models trained from scratch can achieve high performance but often require extensive computational resources and large annotated datasets[15]. By leveraging embeddings from pre-trained models, our approach reduces the need for exhaustive training while still achieving or surpassing the performance of traditional models in certain tasks. The significant improvements over benchmark AUC values in skin lesion classification align with studies emphasizing the effectiveness of deep feature extraction for dermatological applications. The comparable performance in breast lesion classification is consistent with prior findings, although the incremental improvements suggest that embeddings may offer limited advantages in this domain.

The ability to achieve high diagnostic accuracy with embedding-based classifiers has important clinical implications. By reducing computational demands, this approach can facilitate the rapid development and deployment of AI tools in clinical settings, particularly where resources are limited[16]. Improved performance in skin lesion and ocular disease classification can contribute to earlier detection and intervention, potentially enhancing patient outcomes. The use of pre-trained embeddings also promotes the generalizability of models across diverse datasets, which is critical for widespread clinical adoption. The simplicity of linear classifiers makes them more interpretable, aiding clinicians in understanding and trusting AI-assisted diagnosis. Furthermore, the environmental impact of artificial intelligence is becoming recognized. The utilization of high power consuming GPU clusters to perform analysis requires substantial infrastructure and environmental costs[17]. The development of more efficient methods utilizing embeddings as presented herein could provide an approach that is much more efficient, yet comparable in performance.

This study has several limitations. First, the main comparison presented is AUC values. The reference standard utilized for this work only provided AUC values, not other metrics such as precision, recall, and F1 score. Note that we have provided these additional metrics in Table 3. Second, our models were evaluated using publicly available datasets, which, while valuable for research, may not fully represent the diversity and complexity of clinical data encountered in real-world settings. Factors such as varying image quality, differences in imaging protocols, and diverse patient populations may affect the generalizability of our findings. Third, we did not include external validation using independent datasets. The absence of external validation limits the ability to assess the robustness and reproducibility of the models across different clinical environments and patient demographics. Additionally, the linear classifiers employed in this study may not capture intricate non-linear relationships present in medical imaging data, potentially restricting their effectiveness in certain diagnostic tasks. Lastly, the embeddings were generated from models pre-trained on general image datasets rather than specialized medical images. While these models capture broad visual features, they may lack specific representations pertinent to medical imaging, and incorporating domain-specific pre-training could potentially enhance performance. Newly developed models have demonstrated strong capability[18].

## CONCLUSION

In conclusion, our study demonstrates that embedding-based classifiers leveraging pre-trained models like CLIP and ResNet50, combined with simple linear classifiers, can achieve performance comparable to or exceeding that of traditional deep learning models in specific medical imaging classification tasks. This approach offers a computationally efficient alternative that may accelerate the integration of AI technologies into clinical workflows, particularly in resource-constrained environments. Future research should focus on addressing the limitations identified, such as improving performance in complex multi-label tasks, incorporating statistical significance testing, and validating the approach with real-world clinical data. Further exploration into combining embeddings with more sophisticated classifiers may also enhance performance across a broader range of medical imaging applications.

## OTHER INFORMATION

This study utilized publicly available datasets and did not involve the collection of new data from human participants; therefore, it was not registered with a clinical trial registry. While a formal study protocol was not developed for this retrospective analysis, the code and pertinent details for reproducing the experiments will be made available at [insert link to repository]. This research was conducted independently without specific funding or external support, and the authors declare no conflicts of interest.

## SUPPLEMENTARY MATERIAL

*Data Sources*

We utilized five publicly available, de-identified medical imaging datasets to ensure a comprehensive evaluation across diverse medical imaging tasks. All datasets were accessed through publicly available repositories and were used in compliance with their respective data usage agreements.

1. **CBIS-DDSM Breast Cancer Image Dataset:** Comprising approximately 2,864 mammogram images from the Curated Breast Imaging Subset of the Digital Database for Screening Mammography, this dataset is annotated for the presence of benign and malignant lesions, aiding in breast cancer detection.
2. **CheXpert Dataset:** A large dataset of 224,316 chest radiographs annotated for 14 common chest radiographic observations, facilitating evaluation of models in detecting thoracic pathologies.
3. **HAM10000:** This dataset contains 10,015 dermatoscopic images of pigmented skin lesions, categorized into seven diagnostic classes corresponding to common skin conditions and malignancies.
4. **PAD-UFES-20 Dataset:** This dataset includes 2,636 dermatoscopic images of skin lesions, classified as benign or malignant, further supporting skin lesion classification tasks.
5. **Ocular Disease Recognition (ODIR) Dataset:** Consisting of 5,000 retinal fundus images with annotations for various ocular diseases, this dataset enables assessment of models in identifying eye diseases.

*Data Preprocessing*

To prepare the datasets for analysis, we implemented dataset-specific preprocessing steps to standardize the images and facilitate embedding generation. Preprocessing was conducted using the Torchvision library (version 0.19.1+cu121), ensuring compatibility with the pre-trained models and facilitating standardized data handling across datasets. Each dataset was randomly split into a training set (80%) and a testing set (20%), ensuring that individual images were exclusively assigned to one set to prevent data leakage. Missing data was present in two of the datasets. In CheXpert, missing labels were set to zero. For ODIR, missing images were replaced with zero tensors. The remaining datasets did not have missing values. No data augmentation techniques were utilized to maintain consistency with established benchmarks and to focus on evaluating the performance of the embedding-based classifiers without additional enhancements. Image resizing and normalization was performed in a dataset-specific approach as described below:

- **CBIS-DDSM:** Mammogram images were resized and center-cropped to 1,024×1,024 pixels to preserve diagnostic features. Median and Median Absolute Deviation (MAD) normalization were used to account for variations in image brightness and contrast inherent in mammography.
- **CheXpert:** Chest radiographs were resized and center-cropped to 512×512 pixels. ImageNet normalization values were applied. Null values in the labels were imputed with zeros to indicate the absence of specific findings.
- **HAM10000 and PAD-UFES-20:** Images were resized to 224×224 pixels to match the input dimensions expected by the pre-trained models. ImageNet normalization values (mean and standard deviation) were applied to standardize pixel intensity values across images.
- **ODIR:** Retinal images were resized to 224×224 pixels. Median and MAD normalization were applied to accommodate variations in illumination. Missing images were handled by generating zero tensors with appropriate dimensions to maintain consistency in the dataset.

*Demographic and clinical characteristics of cases*

The datasets utilized in this study encompass a diverse range of demographic and clinical characteristics. The HAM10000 (Skin Cancer MNIST) dataset provides information on patient age, sex, lesion location, and diagnosis (7 categories), which will be summarized using descriptive statistics. The CBIS-DDSM dataset includes patient age and breast density, while the Ocular Disease Recognition (ODIR) dataset provides patient age, sex, and the presence of various ocular diseases; these characteristics will be summarized for each data partition (training, validation, testing). The Skin Cancer dataset includes labels for benign and malignant lesions, with the proportion of each class reported for each partition. Finally, the CheXpert dataset may include patient demographics (age, sex) and the presence/absence of 14 chest X-ray observations, which will be summarized for each partition. This comprehensive characterization of the datasets facilitates a nuanced understanding of the patient population and potential confounding factors.

*Predictor and Outcome Variables*

The primary predictor variables in this study were image embeddings generated from pre-trained models, specifically the Contrastive Language-Image Pre-training (CLIP) model and the Residual Neural Network 50 (ResNet50). These embeddings capture high-level features and semantic information from the medical images, effectively summarizing complex visual data into a form suitable for classification tasks. By leveraging pre-trained models, we aimed to utilize rich feature representations without the need for extensive training on each specific dataset, thus enhancing computational efficiency.

The outcome variables differed across the datasets, each corresponding to clinically relevant diagnostic tasks:

1. **CheXpert Dataset**: The outcome variables were binary indicators representing the presence or absence of each of the 14 common chest radiographic observations. These observations included atelectasis, cardiomegaly, consolidation, edema, pleural effusion, pneumonia, pneumothorax, fractures, support devices, absence of findings, enlarged cardiomediastinal, lung lesion, lung opacity, and other pleural abnormalities.

2. **CBIS-DDSM Dataset**: For this mammography dataset, the outcome variable was a binary classification of breast lesions as benign or malignant.

3. **HAM10000 and PAD-UFES-20 Datasets**: These dermatoscopic image datasets involved multiclass labels corresponding to seven diagnostic categories of skin lesions. The categories included actinic keratoses and intraepithelial carcinoma (Bowen's disease), basal cell carcinoma, benign keratosis-like lesions, dermatofibroma, melanoma, melanocytic nevi, and vascular lesions.

4. **Ocular Disease Intelligent Recognition (ODIR) Dataset**: The outcome variables were multi-label classifications of ocular diseases present in each retinal image. Diseases identified included cataract, diabetic retinopathy, glaucoma, myopia, age-related macular degeneration (AMD), hypertension-related changes, and normal findings. Each image could be associated with multiple conditions.

Ground truth labels for all datasets were derived from expert annotations provided by clinicians specialized in the respective fields—radiologists for radiographic images, dermatologists for skin lesion images, and ophthalmologists for retinal images. These expert annotations served as reliable reference standards for evaluating the performance of our classification models. The use of meticulously annotated datasets ensured that the models were assessed against clinically validated outcomes, enhancing the relevance and applicability of our findings.

*Embeddings Generation*

For the first set of embeddings, we utilized a pre-trained ResNet50 model, renowned for its robust performance in image recognition tasks. To adapt ResNet50 for feature extraction, we modified the network by removing its final classification layer. Specifically, we set the model.classifier to an identity function (model.classifier = torch.nn.Identity()), effectively truncating the model to output the activations from its penultimate layer. This adjustment allowed us to extract high-dimensional feature vectors—ResNet50 embeddings—that encapsulate essential visual patterns and structures within each image.

The second set of embeddings was generated using the pre-trained CLIP model, specifically the "openai/clip-vit-base-patch32" checkpoint. CLIP is a multi-modal model trained on a vast dataset of image-text pairs, enabling it to learn visual concepts from natural language supervision. By employing the visual encoder component of CLIP, we extracted embeddings that capture both visual and semantic features of the images. These CLIP embeddings are particularly valuable in medical imaging contexts where subtle visual cues are associated with specific clinical conditions.

*Model Development*

In this study, we developed classification models by leveraging embeddings generated from two pre-trained neural network architectures: the Residual Neural Network 50 (ResNet50) and the Contrastive Language-Image Pre-training (CLIP) model. With the embeddings obtained from ResNet50 and CLIP, we developed classification models using two types of linear classifiers: Logistic Regression (LR) and Support Vector Machines (SVM). These classifiers were chosen for their computational efficiency and suitability for high-dimensional data common in embedding spaces. Linear classifiers were chosen for their simplicity, efficiency, and suitability for data often linearly separable in the embedding space.

<u>*Logistic Regression Models*</u>

Implemented using the Scikit-learn library, Logistic Regression models were configured to handle the variety of classification tasks present in our datasets:

- **Binary Classification**: For datasets like CBIS-DDSM, where the task was to classify lesions as benign or malignant, a binary logistic regression model was employed.
- **Multiclass Classification**: For datasets such as HAM10000 and PAD-UFES-20, involving multiple categories of skin lesions, we configured logistic regression for multiclass classification using a multinomial loss function.
- **Multi-label Classification**: In datasets like CheXpert and ODIR, where images could be associated with multiple conditions simultaneously, we utilized the OneVsRestClassifier wrapper around logistic regression. This approach enabled the model to handle multiple binary classification problems independently for each label, effectively managing the complexity of multi-label scenarios.

*Support Vector Machine Models*

Support Vector Machines were utilized in two implementations using the Scikit-learn library:

- **LinearSVC**: The LinearSVC classifier was employed for its efficiency in handling high-dimensional data, which is characteristic of the embeddings generated by the pre-trained models. It is particularly well-suited for datasets where classes are linearly separable in the embedding space.
- **SVC with Kernel Functions**: For datasets where the relationship between embeddings and class labels might not be strictly linear, we used the SVC classifier with different kernel functions. Both linear and radial basis function (RBF) kernels were applied, depending on the dataset requirements and empirical performance during validation. The RBF kernel, in particular, allows the model to capture non-linear relationships by mapping the input features into a higher-dimensional space.

*Hyperparameter Optimization*

We employed a grid search strategy combined with five-fold cross-validation on the training sets. This approach allowed us to systematically explore a range of hyperparameters and select the optimal configuration for each classifier. For all classifiers, we investigated different values of the regularization strength parameter $C$, which controls the trade-off between achieving a low training error and a low testing error, thus preventing overfitting. The values tested included 0.1, 1, 10, and 100. By adjusting $C$, we aimed to find the balance that yielded the best generalization performance on unseen data.

In the case of the Linear Support Vector Classifier (LinearSVC), we evaluated two loss functions: the standard 'hinge' loss and the 'squared_hinge' loss. The 'hinge' loss is traditionally used in support vector machines and focuses on maximizing the margin between classes, while the 'squared_hinge' loss penalizes misclassifications more severely, potentially enhancing the classifier's robustness to outliers and noisy data. Testing both loss functions allowed us to determine which provided better performance for the specific characteristics of each dataset.

For the Support Vector Classifier (SVC) models applied to the CBIS-DDSM and HAM10000 datasets, we optimized both the kernel function and the gamma parameter. We compared the linear kernel and the radial basis function (RBF) kernel. The linear kernel is appropriate when the data are linearly separable in the embedding space, offering simplicity and efficiency. The RBF kernel, on the other hand, is a powerful tool for capturing non-linear relationships by mapping data into a higher-dimensional space where it may become linearly separable. The gamma parameter, which defines the influence of individual training samples, was tested with 'scale' (which uses $1/(n_{features} \times$ variance of the features) and 'auto' (which uses $1/n_{features}$). Adjusting gamma helped us control the flexibility of the decision boundary, aiming to prevent overfitting while capturing the underlying patterns in the data.

*Model Training*

Across all datasets, we trained a total of 20 distinct models. Each dataset contributed to multiple models by employing both ResNet50 and CLIP embeddings in conjunction with different classifiers. Specifically, for each dataset, we trained independent classifiers using the embeddings generated from ResNet50 and from CLIP, resulting in a comprehensive evaluation of the embedding-classifier combinations. Models were initialized using default settings provided by the Scikit-learn library. All models were trained on standard computing hardware without the need for a graphics processing unit (GPU).

**TABLES**

| Dataset | Total Images | Training Set | Testing Set |
|---|---|---|---|
| MNIST: HAM10000 | 10,015 | 8,012 | 2003 |
| CBIS-DDSM | 3568 | 2864 | 704 |
| Ocular Disease fRecognition (ODIR) | 5,000 | 4,000 | 1279 |
| PAD-UFES-20 | 2298 | 1839 | 459 |
| CheXpert | 223648 | 223414 | 234 |

**Table 1. Datasets Used in the Study and Their Division into Training and Testing Sets.**

| Dataset | Benchmark AUC | Best Embedding Model AUC | Best Embedding Model | Linear Classifier Type |
|---|---|---|---|---|
| MNIST: HAM10000 | 0.609 | 0.935 | Resnet | SVM |
| MNIST: HAM10000 | 0.609 | 0.933 | Resnet | LR |
| MNIST: HAM10000 | 0.609 | 0.9510 | CLIP | SVM |
| MNIST: HAM10000 | 0.609 | 0.9586 | CLIP | LR |
| CBIS-DDSM | 0.464 | 0.5035 | Resnet | SVM |
| CBIS-DDSM | 0.464 | 0.491 | Resnet | LR |
| CBIS-DDSM | 0.464 | 0.4954 | CLIP | SVM |
| CBIS-DDSM | 0.464 | 0.5052 | CLIP | LR |
| ODIR | 0.600 | 0.8544 | Resnet | SVM |
| ODIR | 0.600 | 0.7984 | Resnet | LR |
| ODIR | 0.600 | 0.8577 | CLIP | SVM |
| ODIR | 0.600 | 0.8506 | CLIP | LR |
| PAD-UFES-20 | 0.487 | 0.8576 | Resnet | SVM |
| PAD-UFES-20 | 0.487 | 0.8516 | Resnet | LR |
| PAD-UFES-20 | 0.487 | 0.9120 | CLIP | SVM |
| PAD-UFES-20 | 0.487 | 0.9145 | CLIP | LR |
| Chexpert | 0.723 | 0.7329 | Resnet | SVM |
| Chexpert | 0.723 | 0.7412 | Resnet | LR |
| Chexpert | 0.723 | 0.7462 | CLIP | SVM |
| Chexpert | 0.723 | 0.7490 | CLIP | LR |

**Table 2. Performance of Embedding-Based Classifiers Compared with Benchmark Models Across Medical Imaging Datasets.** Area under the receiver operating characteristic curve (AUC) for each dataset, comparing benchmark models with embedding-based classifiers using ResNet or CLIP embeddings combined with either Support Vector Machines (SVM) or Logistic Regression (LR) classifiers. Across all datasets, the embedding-based classifiers achieved higher AUCs than the benchmark models, indicating improved classification performance. Notably, the highest AUCs were observed with CLIP embeddings and LR classifiers in the MNIST: HAM10000 and PAD-UFES-20 datasets, demonstrating the effectiveness of using pre-trained embeddings for multi-class classification tasks in medical imaging.

| Dataset | Embedding Model | Linear Classifier | Accuracy | Recall | Precision | F1 |
|---|---|---|---|---|---|---|
| MNIST: HAM10000 | Resnet | SVM | 0.7728 | 0.42 | 0.57 | 0.46 |
| MNIST: HAM10000 | Resnet | LR | 0.7798 | 0.5586 | 0.6618 | 0.59 |
| MNIST: HAM10000 | CLIP | SVM | 0.7853 | 0.42 | 0.61 | 0.46 |
| MNIST: HAM10000 | CLIP | LR | 0.8242 | 0.63 | 0.72 | 0.66 |
| CBIS-DDSM | CLIP | SVM | 0.392 | 0.3333 | 0.1307 | 0.1878 |
| CBIS-DDSM | Resnet | LR | 0.402 | 0.353 | 0.3562 | 0.3533 |
| CBIS-DDSM | Resnet | SVM | 0.4261 | 0.3344 | 0.2849 | 0.3066 |
| CBIS-DDSM | CLIP | LR | 0.392 | 0.3333 | 0.1307 | 0.1878 |
| ODIR | Resnet | SVM | 0.0086 | 0.02 | 0.12 | 0.03 |
| ODIR | Resnet | LR | 0.1297 | 0.14 | 0.40 | 0.20 |
| ODIR | CLIP | SVM | 0.0508 | 0.15 | 0.24 | 0.18 |
| ODIR | CLIP | LR | 0.2767 | 0.33 | 0.68 | 0.41 |
| PAD-UFES-20 | Resnet | SVM | 0.5935 | 0.53 | 0.54 | 0.53 |
| PAD-UFES-20 | Resnet | LR | 0.6109 | 0.52 | 0.55 | 0.53 |
| PAD-UFES-20 | CLIP | SVM | 0.6783 | 0.59 | 0.63 | 0.60 |
| PAD-UFES-20 | CLIP | LR | 0.7043 | 0.60 | 0.68 | 0.62 |
| Chexpert | Resnet | SVM | 0.3675 | 0.2331 | 0.6449 | 0.2551 |
| Chexpert | Resnet | LR | 0.3803 | 0.2719 | 0.6166 | 0.3032 |
| Chexpert | CLIP | SVM | 0.4060 | 0.2023 | 0.5799 | 0.2392 |
| Chexpert | CLIP | LR | 0.3974 | 0.1949 | 0.5419 | 0.2325 |

**Table 3. Performance Metrics of Embedding-Based Classifiers Across Medical Imaging Datasets.** Accuracy, recall, precision, and F1-score for embedding-based classifiers using ResNet or CLIP embeddings combined with either Support Vector Machines (SVM) or Logistic Regression (LR) across five medical imaging datasets. In the MNIST: HAM10000 dataset, the highest performance was achieved using CLIP embeddings with LR, yielding an accuracy of 82.42% and an F1-score of 0.66, indicating superior classification of skin lesions. Similar improvements were observed in the PAD-UFES-20 dataset, where CLIP embeddings with LR attained the highest accuracy of 70.43% and an F1-score of 0.62. For the Ocular Disease Recognition (ODIR) dataset, CLIP embeddings with LR also resulted in better performance metrics compared to ResNet embeddings. In contrast, the CBIS-DDSM and CheXpert datasets showed lower overall performance, suggesting that embedding-based classifiers may be less effective for these particular tasks. These findings highlight that combining CLIP embeddings with logistic regression enhances classification performance in certain medical imaging applications.

**FIGURES**

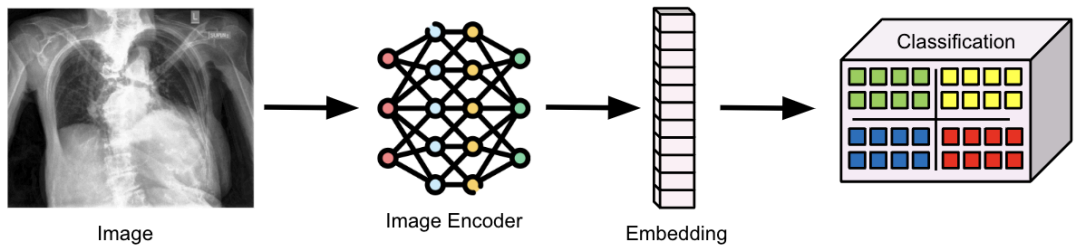

**Figure 1. Workflow for Image Classification Using Pre-trained Embeddings.** Medical images are processed through an image encoder to generate embeddings, which serve as input for classification models. The classification output categorizes images based on predefined tasks, leveraging high-level visual and semantic features extracted from the embeddings.

| Modality/Dataset | Image | Label |
|---|---|---|
| Mammogram<br>CBIS-DDSM | 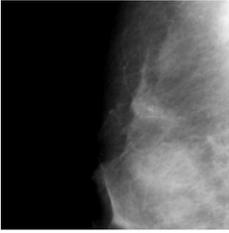 | Benign |
| Chest-X ray<br>Chexpert | 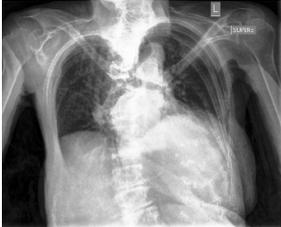 | Normal |
| Retinal Image<br>ODIR | 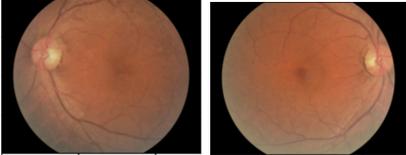 | Normal |
| Skin Lesion<br>PAD-UFES-20 | 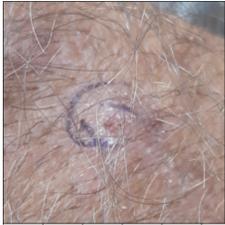 | Basal Cell Carcinoma |
| Skin Lesion<br>HAM10000 | 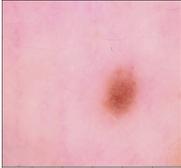 | Melanocytic nevi |

**Figure 2. Example Images Representing Each Class.** Representative images are shown for each modality and classification: mammogram (benign), chest X-ray (normal), retinal image (normal), and skin lesion (basal cell carcinoma and melanocytic nevi).